\def\tdraftstate{0}

\def\tproceedings{1}
\def\tsubmission{2}

\def\doctype{\tsubmission}

\ifnum\doctype=\tproceedings

\documentclass[twoside,leqno,twocolumn]{article}
\usepackage{ltexpprt}

\else

\documentclass[11pt]{article}

\newtheorem{theorem}{Theorem}[section] % section
\newtheorem{lemma}[theorem]{Lemma}

\newtheorem{proposition}[theorem]{Proposition}

\newenvironment{proof}{{\em Proof:}}{\hfill{\hfill\rule{2mm}{2mm}}}

\fi

\usepackage{amsmath}
\usepackage{amssymb}

\def\cF{{\cal F}}
\def\mU{{\mathbb U}}
\def\mM{{\mathbb M}}
\def\G{{\cal G}}
\def\H{{\cal H}}

\def\Vis{\mbox{Vis}}

\def\LVis{\mbox{Lvis}}
\def\mynull{\mbox{NULL}}

\def\Trap{\Gamma}

\mathchardef\lee="0214

\ifnum\tdraftstate=1
\usepackage[notref,notcite]{showkeys}
\fi

\newcommand{\ken}[1]{{%
\ifnum\tdraftstate=1
      {$\rule[-0.12cm]{0.2in}{0.5cm}$\fbox{\tt
            Ken:} }
      #1
      \marginpar{Ken}
      {$\rule[0.1cm]{0.3in}{0.1cm}$\fbox{\tt
            end}$\rule[0.1cm]{0.3in}{0.1cm}$}
\fi %
      }
   }

\newcommand{\kv}[1]{{%
\ifnum\tdraftstate=1
      {$\rule[-0.12cm]{0.2in}{0.5cm}$\fbox{\tt
            KV:} }
      #1
      \marginpar{KV}
      {$\rule[0.1cm]{0.3in}{0.1cm}$\fbox{\tt
            end}$\rule[0.1cm]{0.3in}{0.1cm}$}
\fi %
      }
   }

\begin{document}

%\vspace{-0.6 in}
%
\title{Improved Approximation Algorithms
	for Geometric Set Cover}

\author{Kenneth L. Clarkson\thanks{
         Bell Labs;
         600 Mountain Avenue;
         Murray Hill, New Jersey 07974;
         {\tt clarkson@research.bell-labs.com}}
	\and Kasturi Varadarajan\thanks{University
         of Iowa; {\tt kvaradar@cs.uiowa.edu}}
       }
   
\date{}

\maketitle

\begin{abstract}
Given a collection $S$ of subsets of some set $\mU$, and
$\mM\subset\mU$, the {\em set cover} problem is to find
the smallest subcollection $C\subset S$ such that $\mM$ is
a subset of the union of the sets in $C$.  While the general
problem is NP-hard to solve, even approximately, here we
consider some geometric special cases, where usually
$\mU = \Re^d$.  Extending prior results\cite{BG}, we show that
approximation algorithms with provable performance exist,
under a certain general condition: that for a random subset
$R\subset S$ and function $f()$, there is a decomposition
of the complement $\mU\setminus\cup_{Y\in R} Y$ into an
expected $f(|R|)$ regions, each region of a particular
simple form.  We show that under this condition, a cover
of size $O(f(|C|))$ can be found.  Our proof involves the
generalization of shallow cuttings\cite{m-rph-92} to more
general geometric situations.  We obtain constant-factor
approximation algorithms for covering by unit cubes in
$\Re^3$, for guarding a one-dimensional terrain, and
for covering by similar-sized fat triangles in $\Re^2$.
We also obtain improved approximation guarantees for
fat triangles, of arbitrary size, and for a class of
fat objects.
\end{abstract}

\section{Introduction}
\label{sec:intro}
Given a collection $S$ of subsets of some set $\mU$, and
$\mM\subset\mU$, the {\em set cover} problem is to find
the smallest subcollection $C\subset S$ such that $\mM$ is
a subset of union of the sets in $C$.  In the geometric
setting, almost always $\mU=\Re^d$.
For example, $\mM$ could be a finite set of
points, and $S$ a given finite set of balls. The family
$S$ can be specified implicitly; an example is when $S$ is the set of 
all unit balls. Another interesting example is when $\mM$ 
is the set of points in a simple polygon in $\Re^2$, and
$S$ is the set of visibility regions of the vertices of the
polygon.

The general set cover problem is hard to solve,
even approximately, and the simple greedy algorithm
has performance very close to best possible for a
polynomial-time algorithm, assuming a certain widely
believed complexity theoretic assumption.\cite{F,LY}  Even in 
the geometric setting,
most versions of the problem are believed to be NP-hard,
and indeed NP-hardness has been shown for several
versions. (In some cases, hardness of approximation 
has been shown as well.) The focus
of current work is therefore on obtaining approximation
algorithms that run in polynomial time. Often one obtains
a polynomial-time algorithm guaranteeing a logarithmic
factor approximation by reducing the geometric set
cover problem to the combinatorial set cover problem
\cite{Chvatal,Johnson,Lovasz}.
%; see \cite{Ghosh,?} for some
%applications of this paradigm.

In many cases, the approximation factor can be made
$O(\log c)$, where $c$ is the size of the optimal
solution. Such a result was achieved
for the case of polytope approximation in general
dimension\cite{CPOL}, by applying the {\em iterative reweighting}
approach\cite{littlestone,Welzl,ClaLP} to an associated
set cover problem.  (The reduction of polytope
approximation to set cover was observed by Mitchell and
Suri \cite{ms-sapo-95}.)

Br{\"o}nnimann and Goodrich \cite{BG} showed that a
very similar algorithm applies in the general setting
of set systems with finite VC dimension.\cite{BG} A
key observation of theirs was a connection with {\em
$\epsilon$-nets}.
Consider the subset $\mU_{\epsilon}\subset\mU$,
comprising those points of $\mU$ contained not just in
one set in $S$, but in at least $\epsilon |S|$ of them. %
\kv{Should we replace $\mM$ by $\mU$ here? Is there
a good reason to want the net property only for the 
points we care about?}%
An $\epsilon$-net is a cover for such heavily covered
points.  (That is, the set cover problem is to
find the smallest possible $1/|S|$-net for $\mM=\mU$.)
Suppose that the family $S$ has a $1/r$
net of size $g(r)$, for every $r$ with
$1\le r\le |S|$.
The algorithm of Br{\"o}nnimann and Goodrich
guarantees an approximation factor of
$O(g(c)/c)$, where $c$ is the size $|C|$ of the optimal solution.
For many cases where $g(r) = O(r \log r)$ \cite{C87, hw-ensrq-87},
their algorithm gives an $O(\log c)$ approximation.
Moreover, if $g(r) = O(r)$, such as when $S$ is a family
of disks in $\Re^2$
or halfspaces in $\Re^3$ \cite{msw-hnlls-90, m-rph-92}, they obtain an
$O(1)$ approximation algorithm.

There have been a few other interesting instances where the
$O(\log c)$ factor has been improved upon. Some recent ones
include an $O(\sqrt{\log n})$ approximation factor for covering
an isothetic polygon (with holes) using a minimum number of
rectangles contained in the polygon \cite{ramesh}, and an $O(1)$ 
approximation algorithm for guarding an $x$-monotone polygonal
chain \cite{BKM}.

Hochbaum and Maass \cite{homass} consider the problem of covering
a set of points in the plane with the smallest number of unit disks.
For this and some related problems, they present algorithms, 
that for any $\epsilon > 0$, run in polynomial time and return a 
$(1 + \epsilon)$-approximation.
Since any unit disk may be chosen in the cover, the problem has
a different flavor from that of covering the points using the
minimum number of disks chosen from a set of specified unit disks.

\paragraph{Our Results.}
We generalize results giving small $\epsilon$-nets 
for halfspaces\cite{m-rph-92} to a more general setting\cite{CS}
making a connection between the combinatorial
complexity of the
union of a set of objects and size of a net for the set of objects.
Suppose that $S$ is a set of objects, say triangles in the plane
for concreteness. Suppose that there is a bound $f(j) \geq j$ on
the combinatorial complexity of the boundary of the union of any
$j$ objects from $S$. (More precisely, we need the
number of simple regions in a canonical decomposition of the
exterior of the union of the $j$ objects to be at most $f(j)$.)
We then show, in Theorem~\ref{thm good net},
that for there is a $1/r$ net of size $O(f(r))$,
for every $r\le |S|$.  We apply a ``repair''
or ``alteration'' technique,
using a random sample to divide the problem into
roughly small subproblems, followed by ``repair'' step
in each subproblem.  This approach is similar to Matou{\v s}ek's;
it has been applied also in a similar way to construct
``cuttings.''\cite{c-ochan-91}

As noted, this implies a polynomial time algorithm that
guarantees an $O(f(c)/c)$ approximation factor for covering
a set $\mM$ of points using objects from $S$, where $c$
is the size of the optimal cover.\cite[Theorem 3.2]{BG}
(Note that the result is only interesting for $f(r) = O(r\log n)$;
otherwise the greedy algorithm could be used.)

We give several applications of this result. If $S$ is a
set of fat triangles in the plane, then the combinatorial
complexity of the union of any $j$ elements of $S$
is $O(j \log \log j)$ \cite{MPSSW}, and thus we obtain
$1/r$-nets of size $O(r \log \log r)$ for fat triangles.
This implies, as stated in Theorem~\ref{thm fat tris},
a polynomial algorithm for the corresponding set cover
problem that guarantees an approximation factor of
$O(\log \log c)$. If the triangles in $S$ have roughly the same
diameter, then the union of any $j$ elements from $S$
has a combinatorial complexity of $O(j)$ \cite{MPSSW},
and we obtain $1/r$-nets of size $O(r)$ and an algorithm
for the corresponding set cover problem that guarantees an
$O(1)$ approximation.  There are other applications
in this vein.

Such cover problems are related to wireless network
planning, where the sets in $S$ correspond to antenna coverage
areas.  Prior work has sometimes approximated the coverage
areas as circular disks,\cite{CMWZ} but often such an
idealized model would be far from ideal.  Thus the results
for more general ``fat' objects reported here are relevant.

Another problem that can be viewed as a special case
of wireless network planning is that of guarding
a one-dimensional terrain.  Here, the problem is
to guard the region above an $x$-monotone polygonal chain using the
minimum number of point guards, who are constrained to be on
the chain. The problem was recently studied by Ben-Moshe
et al.~\cite{BKM} who presented a fairly sophisticated
polynomial time algorithm that guarantees an $O(1)$
approximation. We show that a different polynomial-time
constant-factor approximation algorithm can be
derived quite naturally from our paradigm.  The
approximation result is Theorem~\ref{thm guards},
and applies a generalization of the ``Order Claim''
of \cite{BKM} to show, in Lemma~\ref{lem DS},
that an associated sequence
is Davenport-Schinzel.

We next consider the case where $S$ is a set of
axis-parallel unit cubes in $\Re^3$. Boissonat
et al.~\cite{bsty-vdhdc-98} have shown that the
combinatorial complexity of the union of $j$ such cubes is
$O(j)$. Such a bound is however not readily available for
a canonical decomposition of the exterior of the union.
We nevertheless exploit the fact that all the cubes have
roughly the same size to obtain a $1/r$-net of size $O(r)$
and, as stated in Theorem~\ref{thm unit cubes},
a polynomial algorithm for the corresponding set cover
problem that guarantees a factor of $O(1)$.

\section{General results}

\subsection{Small $\epsilon$-nets from small $0$-region sets}
In a geometric setting, the set cover and $\epsilon$-net
problems often have the helpful structure that
for any collection $H\subset S$, the complement
$\mU\setminus\cup(H)$ has a canonical decomposition into
locally defined pieces.  (Here $\cup(H)$ is short-hand for
$\cup_{y\in H} y$.)  That is, there is a set $\cF(S)$
of subsets of $\mU$, such that for any $H\subset S$,
$\mU\setminus \cup(H)$ can be expressed as a union of
sets $\mU\setminus \cup(H) = \cup(\cF_0(H))$, where
$\cF_0(H)\subset \cF(S)$.  Moreover, there is some
integer $b$ so that such decompositions $\cF_0(H)$ can
be described as follows: for each $y\in\cF(S)$, there is
a configuration $B_y\subset S$ of size at most $b$, such
that $y\in\cF_0(H)$ only if $B_y\subset H$ and
$y \cap \cup(H)$ is empty.  Say that $B_y$ {\em defines}
$y$ in that case.  If $y\cap s$ is not empty, for some
$s\in S$, say (as usual) that $s$ meets $y$.  So $y$
is in $\cF_0(H)$ only if no $s\in S$ meets $y$.

It sometimes happens that for some $y\in\cF(S)$ there is
more than one natural configuration $B_y$ that
defines $y$.  To reduce problems with such degenerate
situations, it is often helpful to consider the
regions not only as subsets of $\mU$, but as
configurations $(y, B_y)$, where $B_y$ defines $y$.
Also, the condition that $s\in S$ meets $y$ will
have an analog for configurations, such that $s$
meets or {\em conflicts} with $(y, B_y)$ not only
if $s\cap y$ is nonempty, but also if $s$
takes precedence over a member of $B_y$, for tie-breaking
or other reasons specific to an application.
The set $\cF_0(H)$ will be generalized to comprise
such configurations, and a configuration $(y,B_y)\in\cF_0(H)$
if and only if $B_y\subset H$ and no $s\in H$ conflicts
with $(y, B_y)$, in this broader way.  Even with
this generalization, however, we will have
$\mU\setminus\cup(H)\subset \cup(\cF_0(h))$,
where here $\cup(\cF_0(h)) := \cup_{(y,B_y)\in\cF_0(H))} y$.
We may confuse $(y, B_y)$ with $y$ at times, but
the situation should be clear in context.

We will call the configurations in $\cF_0(H)$ {\em $0$-regions}.
The ``0'' in $\cF_0(H)$ and in $0$-region indicates that the
regions do not conflict with the objects in $H$.
More generally, there
could be $y\in\cF(S)$ that have $B_y\subset R$,
but $(y, B_y)$ conflicts with $j$ members of $H$.
In that case, say that $(y,B_y)\in\cF_j(R)$,
that is, $(y,B_y)$ is a {\em $j$-region} of $R$.  Note that $(y, B_y)$
might be a 0-region with respect to $R$, but a $j$-region
with respect to $S$, that is, conflict with $j$ members of $S$.

Call a given combination of $\mU$, objects $S$, regions $\cF(S)$,
parameter $b$, defining relation, and conflict relation a
{\em configuration system}.  We are assuming that
any point not in $R\subset S$ is in some $0$-region
of $R$.  In such a case, say that the configuration system
is {\em complete}.

This decomposition of the complement puts the problem into
the ``object/region'' framework\cite{CS, CMS},
which is similar to the starter/stopper framework
of Mulmuley.\cite{Mul}
Several properties of the problem
follow from that framework.
A basic property within the framework is the
following version of $\epsilon$-nets,
proven in the objects/regions framework,\cite{C87}
and also in the framework of bounded VC dimension \cite{hw-ensrq-87}.

\begin{lemma}\label{lem likely net}
(Likely $\epsilon$-nets)
For a given complete configuration system,
there is a constant $K$ such that, 
for a random subset $R\subset S$ of size $Kr\log r$,
with probability at least $1-1/r$,
every $0$-region of $R$ is
a $(\lee n/r)$-region with respect to $S$, that is,
a $j$-region with respect to $S$ for some $j\le n/r$.
\end{lemma}

Since our assumption here is that a point not covered
by $R$ is in some $0$-region of $R$, it follows that
$R$ satisfying the condition of the lemma is an $\epsilon$-net,
for $\epsilon = 1/r$ and $|R|\le K r\log r$.
Call an $\epsilon$-net under such conditions a 
{\em likely $\epsilon$-net}. (See
Section \ref{sec:intro} for the definition of an $\epsilon$-net.)
Note that by repeatedly sampling an expected $1 + O(1/r)$ times,
a likely $1/r$-net can be found; also note that an
algorithm for verifying the 
$\epsilon$-net condition would be needed to apply the lemma.

\begin{proof}
See \cite{C87}; also, since the region here
have finite VC-dimension,
the similar results of \cite{hw-ensrq-87} apply.
The proof is simply the union bound, applied to
every $(y, B_y)$; the probability is small
that a particular $j$-region of $S$, with $j\ge n/r$,
is a $0$-region of $R$, and there are $O(n^b)$
$j$-regions.
\end{proof}

We will need the existence of such likely $\epsilon$-nets
under slightly stronger conditions, which
are most conveniently stated simply by
requiring that they exist for any subset of $S$.
\kv{Is it enough to say that $\epsilon$-nets exist
for the objects that meet $\mM'$? }
\ken{doh! yes}

Using the existence of likely $\epsilon$-nets,
and the objects/regions framework, we have the following
generalization of ``shallow cuttings''.

\begin{theorem}\label{thm good net}
For a given complete configuration system, 
let $f(r):= E|\cF_0(R)|$,
where $R\subset S$ is a random subset of size $r$.
Suppose that likely $\epsilon$-nets exist
for any subset of $S$.
Then given $r\ge 2b$, there is a $1/r$-net of size $O(f(r))$.
\end{theorem}

\begin{proof}
The construction is as follows.  Pick a random
subset $R'\subset S$ of size $r$.  For each
$y\in\cF_0(R')$, suppose $y$ meets a set $S'\subset S$,
of size $j'n/r$.  If $j' \le 1$, let let $R_y := \emptyset$;
other, let $R_y$ be a likely $(1/j')$-net
for the objects conflicting with $y$.
Such an $R_y$ will have size at most $K j'\log j'$.
Then $R := R'\cup \cup_{y\in\cF_0(R)} R_y$ is
a $1/r$-net for $S$, by construction.
The expected size of $R$ can be bounded
using Theorem~3.6 of \cite{CS} with $c=2$, and
the ``work'' of that theorem is $K j'\log j'$
for a $(j'n/r)$-region, or no more than
$W(\binom{j}{2})$,
where $j = j' n/r$, and $W()$ is the 
concave ``work'' function 
$W(x) := 4 K \frac{r}{n}\sqrt{x}\log (x\frac{r^2}{n^2})$,
giving a bound
\[
O(W(\frac{n^2}{(r-b)^2} K_{2,b})) f(r)
	= O(f(r)),
\]
assuming $b$ is constant, implying also that the term $K_{2,b}$
of the theorem is constant.
\end{proof}

We note that the proof suggests a natural randomized algorithm to
compute a net. Under appropriate assumptions that certainly
hold for the applications in this paper, the expected running time
of this algorithm is polynomial in the input size.

\subsection{Small covers from small $\epsilon$-nets}

\begin{theorem}\label{thm set cover}
For a given complete configuration system,
with $f(r)$ as in the last theorem,
suppose there is a cover $C\subset S$ of size $c$ for subset
$\mM\subset U$.
Then a cover of $\mM$ of size $O(f(c))$ can
be found in the time proportional to that needed
to construct an $O(1/c)$-net, as in the last theorem, times
a polynomial in~$|S|$.
\end{theorem}

(Note that for particular instances a stronger time bound
can be obtained.)

\kv{Since this a theorem statement, do we need to say that
eps-nets for a weighted version of $S$ implies cover for
$S$? What does `conditions' mean above, the assumption of
previous theorem or its conclusion? One thought is to not
have a theorem here at all -- the reason being that any algorithmic
claim has to backed up in the general setting by oracles for
computing 0-regions, conflicts with these regions, etc.}
\ken{I think the weighting stuff can be subsumed by an
appropriate tie-breaking definition.}

\begin{proof}
The previous theorem implies the existence of
$1/r$-nets of size $O(f(r))$.  This theorem then follows
from Theorem~3.2 of \cite{BG}.   In the algorithm given
to prove their theorem,
$\epsilon$-nets are found many times, for slightly
different sets.  An alternative
approach is to solve the linear programming relaxation,
and find a single $\epsilon$-net.\cite{ERS}
One version of the latter approach is roughly as follows:
solve the linear programming relaxation of the
problem, which yields an assignment, for each object in $s\in S$,
of a value $w_s$ with $0\le w_s\le 1$, such that for
each point $p\in\mM$, it holds that $\sum_{p\in s} w_s\ge 1$.
Then create a multiset $S'$, with "copies" of each $s$,
where the number of copies
is proportional to $w_s$.  Extend the conflict
relation with tie-breaking to allow at most one copy
to contribute to the definition of a region.  The
resulting configuration system has the property that 
every point in $\mM$ is contained in $|S'|/c$
regions; that is, a $1/c$-net is a cover.
\end{proof}

\section{Applications}

\subsection{Covering by Fat Triangles or Regions}

Our first applications of the general results
follow fairly directly from existing combinatorial
bounds and the low complexity of trapezoidal decompositions
in the plane.

\begin{theorem}\label{thm fat tris}
There is a randomized polynomial time algorithm that,
given a set $\mM$ of $m$ points
in $\Re^2$ and a set $S$ of $n$ fat triangles that cover $\mM$,
computes a subset $S' \subseteq S$ of $O(c \log \log c)$ triangles
that cover $\mM$, where $c$ is the size of the smallest subset
of $S$ that covers $\mM$.
\end{theorem}

\begin{proof} (Sketch)
It is long known that the union of $n$ fat triangles
has combinatorial complexity $O(n\log \log n)$. (See \cite{MPSSW},
which also gives a definition of fatness.)
The same bound applies to the canonical
trapezoidal decomposition
of the complement of their union\cite{Mul}; we can
then apply Theorem~\ref{thm good net} with these trapezoids
as the regions.
Similar
remarks apply for fat triangles of approximately the
same size, relying on the sharper bounds known
for  the complexity of their union\cite{MPSSW}
\end{proof}

\begin{theorem}
\label{thm alphabeta}
There is a randomized polynomial time algorithm that,
given a set $\mM$ of $m$ points in $\Re^2$ and a set $S$ of $n$ 
$(\alpha, \beta)$-fat objects of approximately the same size that covers 
$\mM$, computes a subset $S' \subseteq S$ of size $O(\lambda_{s+2}(c))$
that covers $\mM$, where $c$ is the size of the smallest subset
of $S$ that covers $\mM$. Here, $s$ is the maximum number of
intersections between the boundaries of two objects in $S$.
\end{theorem}

Here $\lambda_{s+2}(n)$ is a very-nearly linear function of
$n$, related to the complexity of Davenport-Schinzel sequences.

\begin{proof} (Sketch)
We use a result of Efrat \cite{Efrat} that the combinatorial
complexity of the boundary of the union of $k$ such fat objects
is $O(\lambda_{s+2}(k))$, and proceed as in the case of triangles.
We assume that the trapezoidal
decomposition can be efficiently computed (in polynomial time).
\end{proof}

\subsection{Guarding a Monotone Polygonal Chain}

Let $P$ be a $x$-monotone polygonal chain in $\Re^2$
with $n$ vertices. Let
$\G := \{g_1, \ldots, g_m \}$
be a set of points, which we will call {\em guards},
on $P$.  Say that a guard $g$ lying on polygonal chain $P$
{\em sees} a point $p$ if the line segment $gp$ does not
intersect the region in $\Re^2$ that is strictly below $P$.

Consider the set $\mM_P$ of points in $\Re^2$ that are on
or above $P$.  For $g\in \G$, let
$\Vis(g) := \{p \in \Re^2 | \ g \ \mbox{sees} \ p\}$,
the {\em visibility polygon} of $g$, be the set of
all points seen by $g$.  The problem of guarding $P$
is that of covering the set $\mM_P$ by a small subset of
$S:=\{\Vis(g)\mid g\in \G\}$.  For $S' \subseteq S$, the complement of the 
region covered
by $S'$ is the area between $P$ and the lower envelope of
the visibility polygons in $S'$.  Each point on the $x$-axis has
some corresponding point on the lower envelope (perhaps at
infinity).  It will be helpful, for showing the existence
of a a low-complexity, locally-defined description of the
lower envelope, to consider visibility from the left or
right separately.  It will also be helpful to break ties
among the guards determining the lower envelope at a given
$x$ coordinate.

\subsubsection{Complexity of the Lower Envelope}

Say that $g$ sees $p$ {\em from the left} if $g$ sees
$p$ and $x(g) \leq x(p)$, where $x(p)$ is the $x$-coordinate
of point $p$; define visibility from the right
analogously.  For $g \in \G$, let 
$\LVis(g) := \{p \in \Re^2 | \ g \ \mbox{sees} \ p \ \mbox{from the left}\}$,
the set of points that $g$ sees from the left. Let 
$S_L :=\{\LVis(g)\mid g\in \G\}$.

Fix some subset $\H \subseteq \G$.
Given an $x$-coordinate $x$, say that $g \in \H$
{\em owns $x$ from the left} (relative to $\H$) if there is $y$ such that 
$g$ sees $(x,y)$ from the left, and there is no $y'$, $g' \in \H$
such that $(y', x(g'))$ is lexicographically less than
$(y, x(g))$.  We will say that $g$ owns $x$ from the left
{\em at} $(x,y)$ (relative to $\H$).  If some $x$-coordinate $x$ is owned 
by no point in $\H$, say that $x$ has the owner $\mynull$.

We can now define the {\em ownership diagram} of a set
of guards $\H\subseteq \G$, with respect to $P$; this
definition is for ownership from the left, but similar
definitions and claims apply for ownership
from the right.  The (left) ownership diagram is the
partition of the $x$ axis obtained from the connected
components of each equivalence class of
the relation ``$x$ and $x'$ have the same owner.''
\kv{I introduced connected components here, which is needed.
Please reword if you don't like that sentence.}%
Such components are intervals (or single points),
and so this diagram is a sequence of intervals, each with one
owner.  Call the corresponding sequence of owners,
but excluding $\mynull$, the
{\em ownership sequence} for $\H$.  A key claim for a bound on the
length of this sequence is the following,
a slight generalization of Lemma~2.1 of \cite{BKM}.

\begin{lemma}\label{order}
Suppose $a, b\in \H \subset \G$ and $x, x'\in\Re$ have
$x(a) < x(b) < x < x'$.
Suppose also $a$ owns $x$ (relative to $\H$) at a point $p$, and
$p'=(x', y')$ is seen by $b$.  Then $p'$ is seen by $a$ also.
\end{lemma}

\begin{proof}
Since $a$ owns $x$ at $p$, $a$ sees $p$, and so $P$ is not
above line segment $\overline{ap}$.  Since $b$ is on $P$
and between $a$ and $p$, $b$ in particular is not above
$\overline{ap}$.  Similarly, $P$ is not above segment
$\overline{bp'}$.  Also $p$ is not above $\overline{bp'}$:
if $p$ were above $\overline{bp'}$, it would be seen by
$b$, and since $P$ is not above $\overline{bp'}$, $b$
would also see some point below $p$, but with the same $x$
coordinate, contradicting the assumption that $a$ owns $x$ at $p$.
So $b$ and $P$ are not above $\overline{ap}$, and $p$ and $P$
are not above $\overline{bp'}$.  Therefore $a$ sees $p'$,
as claimed.
\end{proof}

\begin{lemma}\label{lem DS}
An ownership sequence for a set $\H$ of $r$ guards is an $(r,2)$
Davenport-Schinzel sequence, and therefore has length
at most $2r-1$.  It follows that the number of ownership
intervals is no more than $2r$.
\end{lemma}

\begin{proof}
An $(r,2)$ Davenport-Schinzel sequence\cite{SA} is a sequence
of $r$ symbols with no successive entries identical,
and with no subsequence of the form
$a \ldots b \ldots a \ldots b$.
Consider $a , b \in \H$, and first suppose that $x(a) < x(b)$,
as in the previous theorem.
It may be that $a$ owns intervals before $b$ (with smaller
$x$ coordinate than $x(b)$), and it may be that $b$ owns
some intervals to its right, but if $a$ owns some $x$-coordinate
at point
$p$, strictly to the right of $b$, then from the previous lemma,
any point $p'$ with $x(p')>x(p)$ seen by $b$ is also seen
by $a$.  Since $x(a)<x(b)$, such a coordinate would be owned
by $a$ if either $a$ or $b$ owns it, and so could not be
owned by $b$.  Therefore, there is no ownership sequence
of the form $a \ldots b \ldots a \ldots b$.  A similar
argument works if $x(b) < x(a)$, and thus the first
claim of the lemma follows.  The length bound for such
sequences is long-known \cite{SA}.  The final claim follows
because there is at most one interval with owner $\mynull$;
this is the interval to the left of all the guards in $\H$.
\kv{Note this mod which assumes that the chain is strictly
monotone}
\end{proof}

\subsubsection{Guarding in the objects/regions framework}

We will employ Theorem~\ref{thm good net} to compute a
$1/r$-net for the set $S_L$ of size $O(r)$. (Recall that such a
net is a subset $S' \subset S_L$ such that any point belonging
to greater than $|S_L|/r$ sets from $S_L$ also belongs to
some set in $S'$.) In order to apply the theorem, we indicate
explicitly how the configurations and conflicts are defined.
There is a configuration corresponding to every interval
in the ownership diagram for subsets of $\G$ of size
at most 3. Consider an interval $I$ in the ownership
diagram of $\{a,b,c\} \subset \G$, and suppose $b$ owns
each $x \in I$, $a$ owns the interval immediately to
the left of  $I$, and $c$ owns the interval immediately to
the right of $I$\footnote{If $a$ itself owns the interval immediately
to the right of $I$, then such a configuration would
be considered by the subset $\{a,b\}$.}. The set $\{a,b,c\}$
defines this configuration. (The {\em region} of this
configuration is
the set $\{ (x,y) | x \in I, (x,y) \not\in \LVis(b) \}$.) A guard 
$d \in G \setminus \{a,b,c\}$ can conflict with this configuration
in two ways:
\begin{enumerate}
\item Relative to the set $\{a,b,c,d\}$, $d$ rather than $b$ owns
some point $x' \in I$. This of course happens if $d$ sees some point with 
$x$-coordinate $x'$ that lies below the point $p$ at which $b$ owns $x'$ with 
respect to $\{a,b,c\}$. Note that this also happens if $d$ sees $p$
and $x(d) < x(b)$.
\item Relative to the set $\{a,b,c,d\}$, $b$ continues to own
all points in $I$ but the interval immediately to the left of
$I$ is owned by $d$ and not $a$. Because of the way we break
ties in defining ownership, this is not a pathological situation
at all. A conflict also occurs if $d$ owns the interval
immediately to the right of $I$ in the ownership diagram
of $\{a,b,c,d\}$.
\end{enumerate}
  
%Consider now the coarsest partition respecting the
%left ownership diagram and the right ownership diagram.
%(Such a partition results from merging the sorted lists
%of interval endpoints.)  The number of resulting intervals
%is no more than $8r$.  Each open interval $I$ is owned by
%from the left by some guard $b$ in $G$ or null, and from
%the right by some guard $b'$ in $G$ or null.  Similar
%there is some $a$ owning the left endpoint, and some $c$
%owning the right endpoint.  Moreover, $I$ is also in the
%merged ownership diagram of $\{a,b,b',c\}$, with the same
%ownerships.  It is also clear when a guard $g$ conflicts
%with $I$ (or rather, with the given configuration).
%Some such conflicts are due to the visibility polygon
%of $g$ meeting region between the lower envelope and $P$
%in the interval $I$.  A guard could also conflict if the
%lower envelope with $g$ were the same, but $x(g)<x(a)$;
%that is, a conflict for the configurations
%corresponds to the tie-breaking for ownership.

With these definitions, observe that the size of $\cF_0(\H)$,
for any subset $\H \subset \G$, is exactly equal to the
number of intervals in the ownership diagram of $\H$, which
is $O(|\H|)$ by Lemma~\ref{lem DS}. We can therefore
use the algorithm of Theorem~\ref{thm good net} to compute
in randomized polynomial time a $1/r$ net for $S_L$ of
size $O(r)$. We define $S_R$ in a manner symmetric to $S_L$,
and note that the union of a $1/2r$ net for $S_L$ and a
$1/2r$ net for $S_R$ is a $1/r$ net for $S$.

The above arguments are readily adapted to the case
where there can be multiple copies of each guard.
We can therefore use apply Theorem~\ref{thm set cover}, and so a set 
of guards for $P$ can be found with
a polynomial time algorithm, of size within a constant factor
of optimal.

\begin{theorem}\label{thm guards}
Let $P$ be a $x$-monotone polygonal chain in $\Re^2$
with $n$ vertices. Let
$\G := \{g_1, \ldots, g_m \}\subset \Re^2$ be guards,
such that $\mM_P$ is seen by $\G$.  Then a subset $C\subset \G$
that also sees $\mM_P$, of size within $O(1)$ of optimal,
can be found in polynomial time.
\end{theorem}

\subsection{Covering with Cubes}

We now consider the set cover problem where $\mM$ is a set of
$m$ points in $\Re^3$ and
$S$ is a set of $n$ axis-parallel unit cubes in $\Re^3$ that cover
$\mM$. We first show that any $1 \leq r \leq n$, there is a $1/r$-net for
$S$ of size $O(r)$. That is, there is a subset $T \subseteq S$
with $|T| = O(r)$ such that any point that
is contained in at least $n/r$ cubes from $S$ is also contained
in some cube from $T$. We also present a randomized polynomial
time algorithm to compute such a $1/r$-net. From Lemma~\ref{lem likely net},
it is possible to compute a $1/r$-net
of size $O(r \log r)$ in randomized polynomial time. 

Let $G$ be the vertices of a grid in $\Re^3$ of side $1/2$. That is,
\[ G := \{(\frac{i}{2}, \frac{j}{2}, \frac{k}{2}) \ | \ i, j, k,
        \ \mbox{are integers} \ \}.\]

We ``assign'' each cube $C \in S$ to some point in $G$ that lies
in the interior of $C$. (Note that there is always at least one
such point.) Let $S[p] \subseteq S$ denote the set of cubes
assigned to the point $p \in G$. For each $p \in G$ such that
$|S[p]| \geq \frac{n}{dr}$, where $d > 0$ is a suitably large constant,
we compute a $\frac{n}{dr |S[p]|}$-net $T[p]$ for $S[p]$ 
of size $O(\frac{|S[p]| dr}{n})$
using the procedure described below.  Let 
\[ T := \bigcup_{p \in G; |S[p]| \geq \frac{n}{dr}} T[p]. \]

Clearly,
\[ |T| \leq \sum_{p \in G} O(\frac{|S[p]| dr}{n}) = O(dr). \]
We argue that $T$ is a $1/r$-net for $S$. Let $q \in \Re^3$ be any
point that is covered by at least $n/r$ cubes from $S$. Consider
the cube $E$ of side length 2 that is centered at $q$. Each cube in $S$ 
that contains $q$ is contained in $E$, so it must have been assigned to one 
of the at most $d$ points in $G \cap E$. It follows that there is a point
$p \in G \cap E$ such that $S[p]$ has at least $\frac{n}{dr}$ cubes
that contain $q$. Thus $T[p]$, and hence $T$, will have a cube
that contains $q$.

\paragraph{A net for a cluster.} We now describe a randomized
polynomial time algorithm for computing a $1/r$-net, for any
$1 \leq r \leq |S[p]|$, for a ``cluster'' $S[p]$. The special
property of $S[p]$ is that there is a point, namely $p$, that
lies in the interior of all the cubes in $S[p]$. For any
non-empty subset $S' \subseteq S[p]$, we define a canonical
trapezoidation of the boundary of the union of the cubes
in $S'$. This is obtained by taking, for each face of each cube
in $S'$, a canonical trapezoidation of the (isothetic polygon 
corresponding to the) portion of the face
that lies on the boundary of the union of $S'$. Let $\Trap(S')$
denote the canonical set of trapezoids thus obtained.

\begin{proposition}
For any subset $S' \subseteq S[p]$, $|\Trap(S')| = O(|S'|)$.
\label{trap-decomp}
\end{proposition}

\begin{proof}
Boissonat et al.~\cite{bsty-vdhdc-98} show that the combinatorial complexity
of the boundary of the union of cubes in $S'$ is $O(|S'|)$. The proposition
follows because $\Trap(S')$ is linearly bounded by the
combinatorial complexity of the boundary of the union of $S'$.
\end{proof}

We define the ``region'' $\mu_{\tau}$ corresponding to the trapezoid
$\tau \in \Trap(S')$ to be the set of all points $q \in \Re^3$ for
which $\tau$ intersects the segment $qp$ in the relative
interior of the segment. It is easy to see, using the fact that
$p$ lies in the interior of all the cubes in $S[p]$, that
the regions $\{ \mu_{\tau} \ | \tau \in \Trap(S') \}$ partition
the exterior of the union of the cubes in $S'$. The sets that
define and conflict with a region $\mu_{\tau}$ are defined in
the standard way: a cube $C \in S[p]$ will conflict with
$\mu_{\tau}$ if $C$ contains a point in $\mu_{\tau}$. 
We can therefore apply Theorem~\ref{thm good net} to compute
a $1/r$ net for $S[p]$ of size $O(r)$.
 
Putting everything together, we have the following:

\begin{lemma}
There is a randomized polynomial time algorithm that, given
a set $S$ of $n$ axis-parallel unit cubes in $\Re^3$,
and a parameter $1 \leq r \leq n$, computes a subset
$T \subseteq S$ of $O(r)$ cubes with the property that
any point that is contained in at least $n/r$ cubes in $S$
is contained in some cube from $T$.
\end{lemma}

It is also straightforward to handle the case where there
can be multiple copies of each cube. Plugging this lemma into the 
approach of Theorem~\ref{thm set cover}, we
obtain the following result for the corresponding geometric set covering 
problem.  

\begin{theorem}\label{thm unit cubes}
There is a randomized polynomial time algorithm that, given
a set $\mM \subseteq \Re^3$ of $m$ points and a
a set $S$ of $n$ axis-parallel unit cubes in $\Re^3$ that cover
$\mM$, computes a subset $T \subseteq S$ of $O(c)$ cubes 
that cover $\mM$, where $c$ is the size of the smallest subset
of cubes from $S$ that covers $\mM$.
\end{theorem}

We remark that the problem of covering a given set of points by the
smallest number of axis-parallel unit cubes, where we are allowed 
to pick any axis-parallel unit cube in our cover, admits a polynomial
time approximation scheme \cite{homass}. 

\section{Conclusion}
It is worth exploring other versions of the geometric set cover
problem where better approximation guarantees can be obtained
via improved bounds on $\epsilon$-nets. Our work also highlights
the need for a deeper understanding of the connection between bounds 
on the union and the size of $\epsilon$ nets.  

We close with a natural open problem, which is to obtain polynomial-time
approximation algorithms with a sub-logarithmic guarantee for
the geometric set cover problem where $\mM$ is a set of $m$ points in 
$\Re^3$, and $S$ is a set of $n$ unit balls whose union covers $\mM$.

{\bf Acknowledgement.} The authors appreciate many helpful discussions
with Chandra Chekuri.

\bibliographystyle{alpha}
\bibliography{p}

\end{document}